\documentclass{article}

\usepackage{arxiv}
\usepackage{graphicx}
\usepackage[utf8]{inputenc} 
\usepackage[T1]{fontenc}    
\usepackage{hyperref}       
\usepackage{url}            
\usepackage{booktabs}       
\usepackage{amsfonts}       
\usepackage{nicefrac}       
\usepackage{microtype}      
\usepackage{lipsum}

\title{Search for Dark Sector Physics at the NA64 experiment
in the context of the Physics Beyond Colliders Project}

\author{
  Dipanwita Banerjee\thanks{Speaker for the XXIX International Symposium on Lepton Photon Interactions at High Energies - August 5-10, 2019, Toronto, Canada} \\
  CERN/University of Illinois\\
  On behalf of the NA64 Collaboration and the Physics Beyond Colliders Conventional Beams Working Group\\
  \texttt{dipanwita.banerjee@cern.ch} \\
}

\begin{document}
\maketitle

\begin{abstract}
The NA64 electron beam program comprises of a high sensitivity search for visible and invisible decays of the hypothetical dark photon, $A'$, with a goal to either observe the sub-GeV Dark Matter mediator or exclude most of the model parameter space. The visible channel search also includes clarification of the origin of the $^8$Be anomaly observed by the Atomki experiment. The NA64 collaboration further aims to expand its searches with a proposal to use the muon beam at the CERN M2 beam line which will focus on the unique possibility to search for a new scalar or vector states weakly coupled predominantly to muons, e.g. a new Z$_\mu$ gauge boson of L$_\mu$ - L$_\tau$ symmetry, which might explain the long standing muon (g-2)$_\mu$ anomaly. It will also include searches for the Z$_\mu$ as a vector
mediator of Dark Matter production. Within the Conventional Beam Working Group of
the Physics Beyond Colliders (PBC) study, several projects for the muon beamline (M2) in the CERN North Area were proposed. The different experimental requirements and the various technical feasibility studies performed by the group will be presented together with the combined results of NA64 electron beam run between 2016-2018, its future plans and the muon proposal with its planned searches.
\end{abstract}

\section{Introduction}
Dark Matter still remains a great puzzle despite the intensive searches and efforts at the LHC and non-accelerator experiments \cite{ref1}. Though various Dark Matter (DM) models have been ruled out by stringent constraints obtained on DM coupling to Standard Model (SM) particles, little is known about the origin and dynamics of the dark sector itself. Several models of DM suggest the existence of dark sectors consisting of SU(3)$_C$ $\times$ SU(2)$_L$ $\times$ U(1)$_Y$ singlet fields. The difficulty in the searches arise because these sectors of particles do not interact with ordinary matter directly but could couple to it via gravity. In addition to gravity, however, there might be another very weak interaction between ordinary and dark matter mediated by $U'(1)$ gauge bosons, $A'$ (dark photons), mixing with SM photons with coupling strength, $\epsilon$, where $\epsilon$ $<<$ 1. Review is  available in \cite{ref3}.
\paragraph{}
NA64 is an experiment proposed in December 2013, which aims to explore this dark sector and discover or potentially rule out a considerable parameter space region for sub-GeV dark matter models by looking for the dark photon, $A'$, which is the force carrier of the dark sector. The dark photon, depending on its mass, can decay to DM particles, via an invisible channel, if M$_{A'}$ $\geq$ 2M$_{\chi}$, where M$_{A'}$ is the mass of $A'$ and M$_{\chi}$ is the mass of the dark matter particles, or to standard model particles, $e^+e^-$, via a visible channel, when M$_{A'}$ $<$ 2M$_{\chi}$. NA64 was approved in February 2016 and had regular runs since, until the Long Shutdown 2 at CERN in 2018. The latest results of NA64 for both channels will be presented along with the plans to extend the search to muon beams.
\section{Invisible Decay Channel $A'$ $\rightarrow$ $\chi \chi$}
\begin{figure}[h]
\centering
\begin{minipage}{0.45\textwidth}
\includegraphics[scale=0.23]{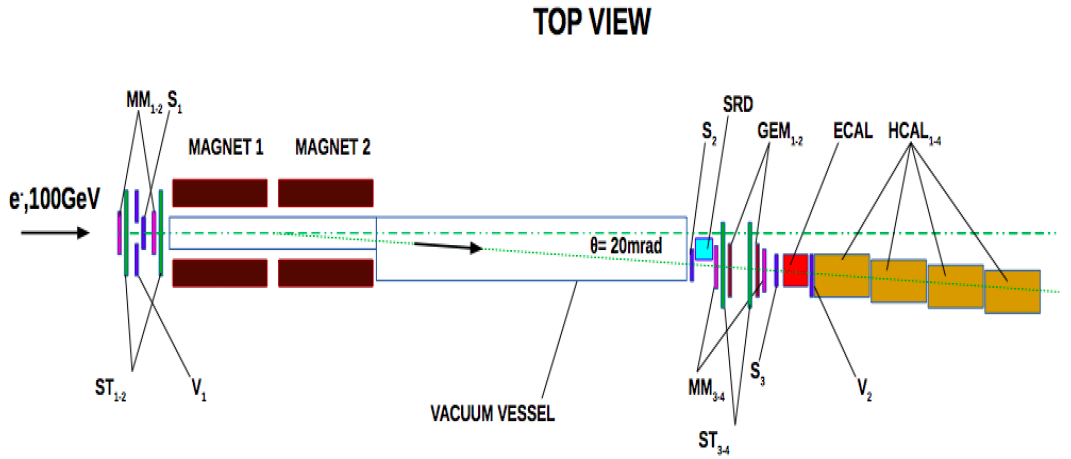}
\caption{Setup of NA64 for the invisible decay channel search of $A'$ $\rightarrow$ $\chi \chi$ \cite{na64_paper1}.}
\label{fig1}
\end{minipage}
\hfill
\begin{minipage}{0.45\textwidth}
\centering
\includegraphics[scale=0.25]{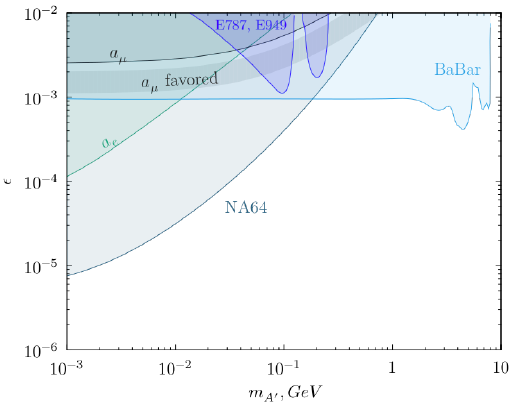}
\caption{The NA64 90 $\%$ C.L. exclusion region in the (M$_{A'}$, $\epsilon$) plane \cite{na64_mk}.}
\label{fig8}
\end{minipage}
\end{figure}

\begin{figure}[h]
\centering
\includegraphics[scale=0.25]{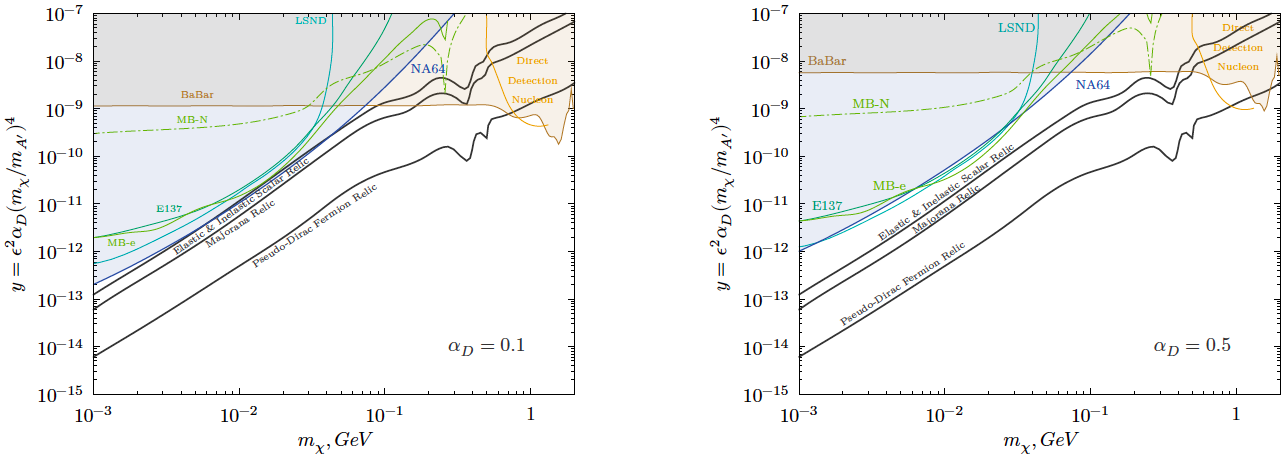}
\caption{The NA64 sensitivity limits in the (M$_\chi$, $y$) plane along-with the limits from other experiments \cite{na64_mk}.}
\label{fig9}
\end{figure}

NA64 is a fixed target experiment searching for dark photons from the interaction of electrons dumped in an active target (ECAL). Detailed setup of the experiment is shown in Fig \ref{fig1}. The experiment employs the 100 GeV/c momentum electron beam from the H4 beamline at the CERN SPS, at a maximal beam intensity of about 10$^7$ e$^-$ per 4.8 s spill. A fraction of the incoming beam energy of about 0.5$E_0$ is deposited in the ECAL, while the rest is carried by the light DM particles, $\chi$, from the prompt $A'$ $\rightarrow$ $\chi\chi$ invisible decay. As $\chi$'s interact weakly with ordinary matter, the $A'$ $\rightarrow$ $\chi \chi$ decay  signature in the experiment would be the missing energy of more than 0.5$E_0$. This requires precise knowledge of the incoming beam and a completely hermetic detector.
All detector responses were simulated and validated with data as best possible to determine the signal cuts and minimise the background with maximizing the efficiency of $A'$ selection. Details of the detectors and their simulations are included in \cite{na64_paper1}. Additionally, in order to validate the simulation of $A'$ production and the sensitivity of the search, the rare QED process of dimuon production was used as a reference process. Simulation of this process and its validation with data was used to set a conservative estimate of about 10$\%$ for the systematic uncertainty on the $A'$ yield which was the dominant source of error. Following this the overall detection efficiency of 0.5-0.6 was reached depending on the mass of $A'$.
\paragraph{}
No signal was observed in the combined data sample of 2016-2017 of about 10$^{11}$ electrons on target (EOT) with a combined background of 0.2 events. Data analysis of the 2018 run with 1.9 $\times$ 10$^{11}$ EOT is ongoing, however the combined sensitivity on the total 3 $\times$ 10$^{11}$ EOT has already been calculated as shown in Fig \ref{fig8} for the $m_{A'} - \epsilon$ parameter space. The dark matter sensitivity in the $y - m_{\chi}$ parameter space where $y = \epsilon^2 \alpha_D (m_{\chi}/m_{A'})^4$, with $\alpha_D$ being the $\chi$-$A'$ coupling constant, is also shown in Fig \ref{fig9}. The advantage of NA64 is that its sensitivity  is proportional to $\epsilon^2$ as opposed to other beam dump experiments where  the sensitivity is proportional to $\epsilon^4$. Thus, with 3 $\times$ 10$^{11}$ EOT and $\alpha_D$ = 0.1 NA64 is the first beam dump experiment to have touched the Elastic and Inelastic Scalar Relic sensitivity.
 \section{Visible Decay Channel $A'$ , X $\rightarrow$ $e^+e^-$}
\begin{figure}[h]
\centering
\begin{minipage}{0.45\textwidth}
\centering
\includegraphics[scale=0.22]{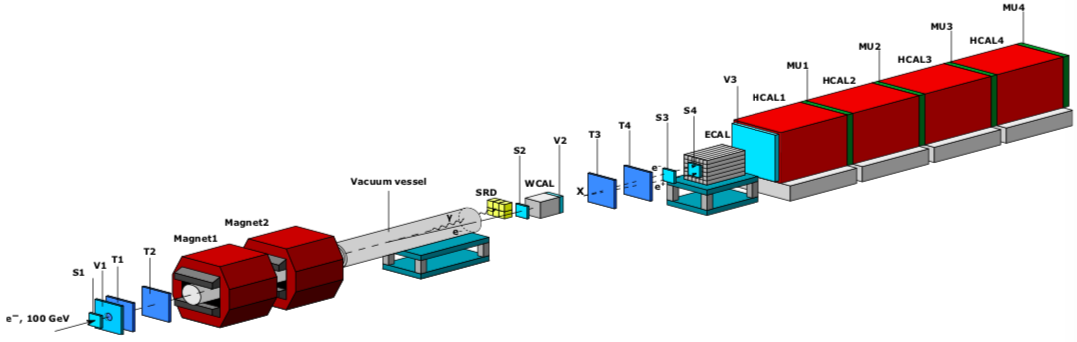}
\caption{Setup of NA64 for the visible decay channel search of $A'$ , X(17 MeV) $\rightarrow$ $e^+e^-$ \cite{na64_paper2}.}
\label{fig10}
\end{minipage}
\hfill
\begin{minipage}{0.45\textwidth}
\centering
\includegraphics[scale=0.3]{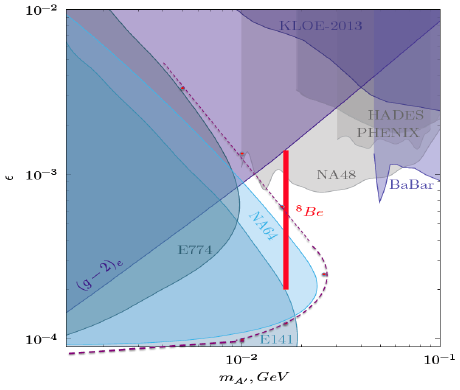}
\caption{ The parameter space for $A'$, X $\rightarrow$ $e^+e^-$ excluded by the NA64 experiment with the 2017 data (blue shadowed region) and with 2017+2018 data (dashed line). The red vertical line is the region that could explain the $^8$Be anomaly \cite{na64_mk}.}
\label{fig11}
\end{minipage}
\end{figure}

In addition to searching for invisible decays of $A'$, NA64 is also equipped to search for the the visible decays of $A'$ to Standard Model particles. One strong motivation for the visible channel comes from the $^8$Be anomaly observed by the ATOMKI group \cite{atomki}, wherein they detected a bump in the invariant mass distribution of the $e^+e^-$ pair coming from the nuclear transition of the excited Be* atom to its ground state. This bump can be explained by a 17 MeV X boson assuming a non-universal coupling to quarks and a coupling to electrons in the range 2 $\times$ 10$^{-4}$ $<$ $\epsilon$ $<$ 1.4 $\times$ 10$^{-3}$. The $A'$ explanation of this anomaly has already been excluded from the $\pi^0$ $\rightarrow$ X$\gamma$ decay \cite{na48}. In order to perform this search NA64 adds another active Tungsten target, the WCAL, as shown in Fig \ref{fig10}, made as short as possible to maximise the sensitivity to short lifetimes while keeping the leakage of particles at a small level. The criteria for the event selection is: (i) No energy deposition in the V2 above the zero energy threshold; (ii) The signal in S4 consistent with two MIPs; (iii) Sum of energies in WCAL+ECAL = E0 (beam energy); (iv) The lateral and longitudinal shape of the ECAL shower consistent with a single electromagnetic one; (v) The maximal energy deposition in the central ECAL cell.
From 2017 to 2018 in order to improve the sensitivity to short lived X bosons the beam momentum was increased from 100 GeV/c to 150 GeV/c for the visible channel, thinner V2 counters were installed immediately after the WCAL, a vacuum pipe was added between the WCAL and ECAL and the distance between them was increased. The dominant background for this search comes from the $K_{S}^0$ $\rightarrow$ $\pi^+\pi^-$ decay from the $K^0$s produced by hadrons misidentified as electrons. In order to estimate the level of this background and the flux of $K^0_S$, both data and simulation were used. Both methods gave comparable number for the $K^0_S$s and the number of background events thus predicted for the 2017 run was 0.06 and that for the 2018 run was 0.006. The improvement in 2018 is expected as the distance between the WCAL and ECAL was increased and therefore less $K_0$ events pass the criteria of having the maximal energy deposition in the central ECAL cell. Further details of the search are presented in \cite{na64_paper2}. The sensitivity was calculated for the combined data of the 2017 and 2018 run as shown in Fig \ref{fig11} with the 17 MeV band for the $^8$Be anomaly.
\section{Future plans for NA64$\mu$}
\begin{figure}[h]
    \centering
    \begin{minipage}{0.45\textwidth}
    \centering
    \includegraphics[scale=0.22]{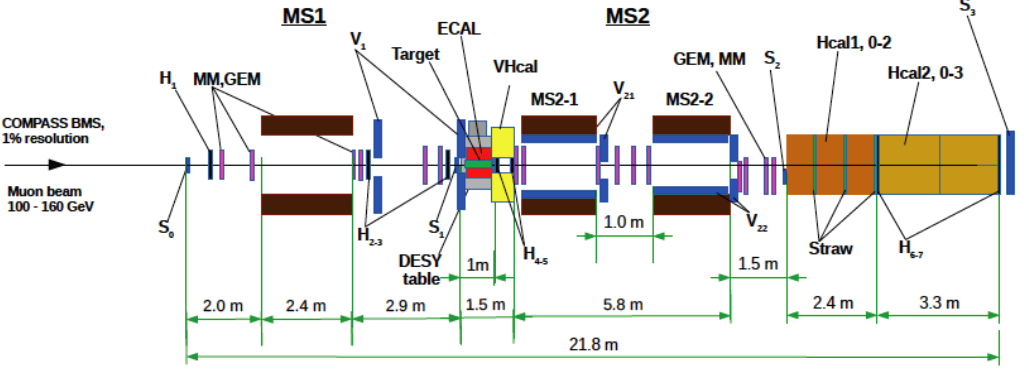}
    \caption{Schematic of the NA64 experimental setup for the Muon Run \cite{na64mu}.}
    \label{fig12}
    \end{minipage}
    \hfill
    \begin{minipage}{0.45\textwidth}
    \centering
     \includegraphics[scale=0.17]{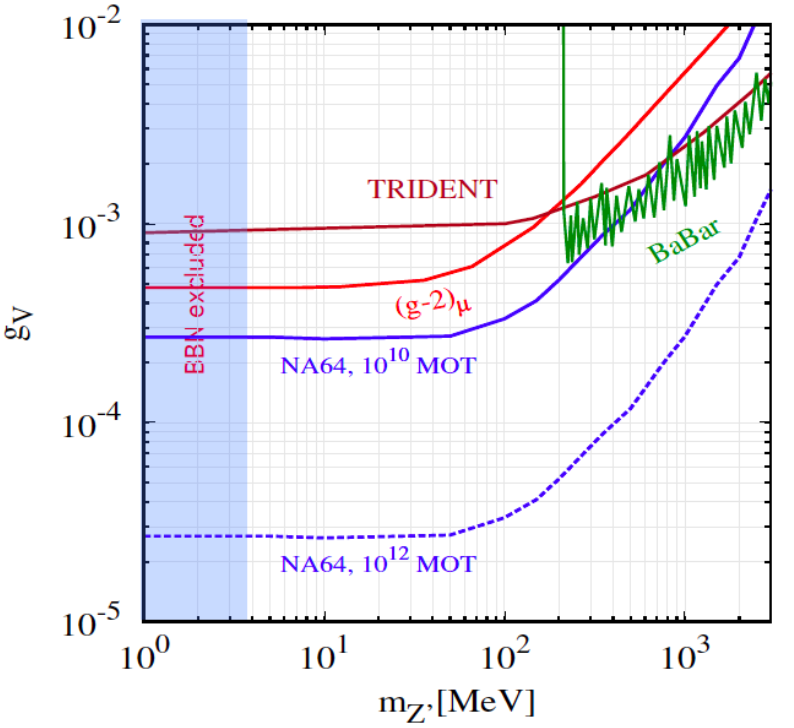}
    \caption{The NA64$\mu$ projected bounds calculated for 10$^{10}$ and 10$^{12}$ muons on target \cite{na64mu}.}
    \label{fig13}
    \end{minipage}
\end{figure}

\begin{figure}[h]
    \centering
    \includegraphics[scale=0.22]{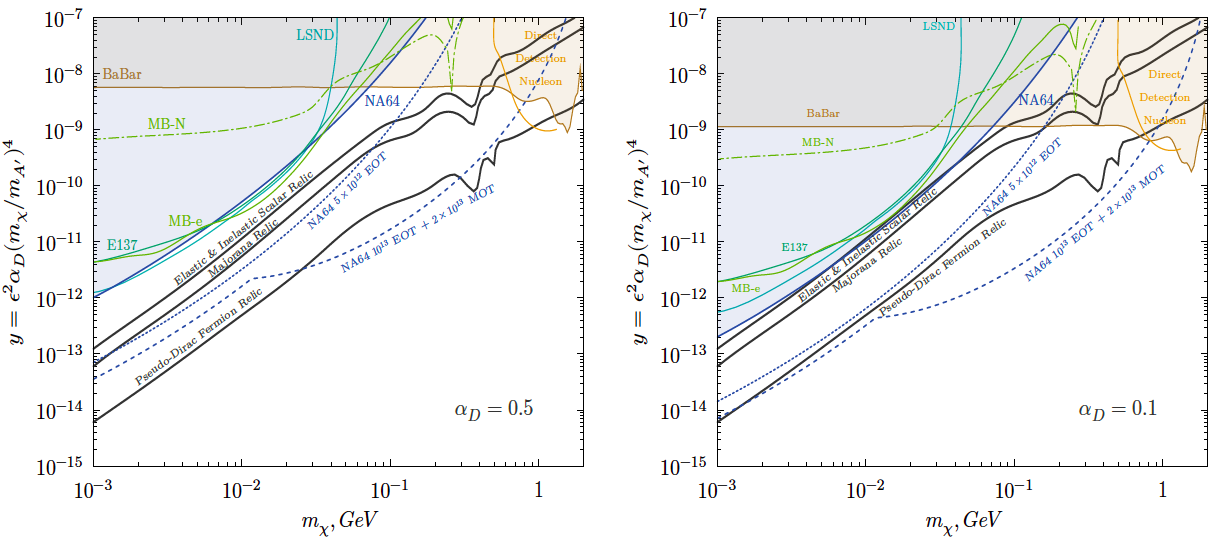}
    \caption{The NA64projected bounds calculated for 5 $\times$ 10$^{12}$ EOT and the combined sensitivity of 10$^{13}$ EOT + 2 $\times$ 10$^{13}$ MOT  \cite{ldm}.}
    \label{fig14}
\end{figure}

NA64 now plans to extend its search with muon beams, also known as NA64$\mu$, at CERN which will broaden the potential of the experiment with the setup shown in Fig \ref{fig12}. A dark sector of particles predominantly weakly-coupled to the 2nd and possibly 3rd generations of the standard model is motivated by several theoretically interesting models \cite{g-2A}. Additional to gravity this new very weak interaction between the visible and dark sector could be mediated either by a scalar $S$ or $U'(1)$ gauge bosons ($Z_{\mu}$) interacting with ordinary muons. If a light $Z_{\mu}$ mediator, with the coupling strength lying in the experimentally accessible region exists, it could also explain the (g-2)$_{\mu}$ anomaly \cite{g-2} - the discrepancy between the predicted and measured values of the muon anomalous magnetic moment. Since the mass of the $Z_\mu$ is currently constrained to less than 200 MeV NA64$\mu$ focuses on the invisible decays to neutrinos or light Dark Matter particles. 
\paragraph{}
The principle of the experiment is to employ the high energy muon beam from the CERN SPS M2 beamline \cite{M2} of momentum 160 GeV/c and dump it against the target (ECAL) \cite{g-2A}. The incoming momentum of the beam is tagged by the spectrometer MS1. In the event of production of $Z_{\mu}$ the scattered muon carries a fraction of the momentum, e.g., less than 0.5$E_0$, while the rest is carried by the $Z_{\mu}$ which subsequently decays to neutrinos or light DM particles. The momentum of the scattered muon is reconstructed by the spectrometers, MS2-1 and MS2-2, and as the neutrinos or the dark matter particles will not interact with the detectors downstream, a missing momentum gives the signature of the search. Because of this, suppressing the background from the low momentum tail of the beam as well as precise reconstruction of the momentum of the scattered muon is extremely crucial. This requires very good knowledge of the beam and preparation of beam parameters suited for optimal performance of the detectors. These studies were performed by the PBC Conventional Beams Working Group. As there are also other proposals in addition to NA64 for the M2 beamline, including the successor to the COMPASS experiment and MUonE, aiming to investigate the hadronic contribution to the vacuum polarisation, feasibility studies have also been performed to check the location and the potential to run more than one experiment in parallel. Given the acceptance of the NA64 detectors a parallel beam of size 10 cm diameter and good momentum definition is desirable. The studies showed that the required beam paramteres for the proposed location are feasible which were used for the detector simulation and estimation of their performance. Detailed report of the studies and the results are enclosed in \cite{PBC1} \cite{PBC2}. In addition to having precise knowledge of the beam, ensuring a completely hermetic detector and suppression of background from beam contamination is also essential. Detailed simulations have been performed to estimate the sensitivity of the search \cite{na64mu} and the projection for the parameter space and dark matter search is shown in Figs \ref{fig13} and \ref{fig14} respectively. As seen for 10$^{10}$ muons on target (MOT) which is possible in 2 weeks of data taking in the high intensity beam, NA64$\mu$ will be able to cover the (g-2)$_{\mu}$ favored parameter space and combining the electron and muon runs it is able to probe a considerable region of the parameter space.
\section{Conclusion}
The prospects of search for sub-GeV Dark Matter are presented in the scope of NA64 that combines beam dump experiments with missing energy/momentum techniques. With the accumulated statistics between 2016-2018 NA64 is already able to probe a considerable parameter space in the sub-GeV mass range for the visible and invisible channels and is the first beam dump experiment to touch the Elastic and Inelastic Scalar Relic sensitivity for $\alpha_D$ = 0.1. In addition, with the visible channel search NA64 also probed a portion of the parameter space favoured by the X boson explanation of the Be-anomaly. In 2021 NA64 plans to accumulate about 10$^{12}$ EOT for the invisible channel and around 5 $\times$ 10$^{11}$ EOT for the visible channel to probe new areas of the parameter space. It also plans to setup the experiment in the M2 beamline in 2022 and have its first muon beam run to probe the Z$_{\mu}$ parameter space. The studies have been performed in collaboration with the PBC Conventional Beams Working Group and the sensitivity of the search has been estimated. With about 10$^{10}$ MOT NA64$\mu$ will constrain the (g-2)$_{\mu}$ favoured parameter space. As presented with 5 $\times$ 10$^{12}$ EOT NA64 will be able to test the Scalar and Majorana scenarios for $\alpha_D$ = 0.1, while with a combined run with muons of 10$^{13}$ EOT + 2 $\times$ 10$^{13}$ MOT NA64 has the potential to fully explore the parameter space of Light Dark Matter \cite{ldm}.

\end{document}